\documentclass[a4paper,fleqn,usenatbib]{mnras}

\usepackage[T1]{fontenc}
\usepackage{ae,aecompl}

\usepackage{graphicx}	% Including figure files
\usepackage{amsmath}	% Advanced maths commands
\usepackage{amssymb}	% Extra maths symbols
\usepackage{multicol}        % Multi-column entries in tables
\usepackage{bm}		% Bold maths symbols, including upright Greek
\usepackage{color}
\usepackage{xcolor}

\title[{\em TESS} observations of Be stars]
{{\em Tess} observations of Be stars:
general characteristics and the impulsive magnetic rotator model}
\author[L. A. Balona and D. Ozuyar]{\
L. A. Balona$^{1}$\thanks{E-mail: lab@saao.ac.za} and
D. Ozuyar$^{2}$
\\
$^1$ South African Astronomical Observatory, 
P.O. Box 9, Observatory 7935, South Africa\\
$^2$ Ankara University, Faculty of Science, 
Dept. of Astronomy and Space Sciences, 06100, Tandogan, Ankara, Turkey}

\begin{document}

\date{Accepted .... Received ...}

\pagerange{\pageref{firstpage}--\pageref{lastpage}} \pubyear{2011}

\maketitle

\label{firstpage}

\begin{abstract}
In this work, we study the general characteristics of Be star light curves 
from {\em TESS} sectors 1--26 in the light of a model we call the ``impulsive 
magnetic rotator''.  It is found that 73\,percent of 441 classical  Be stars 
are short-period variables. The light curves are characterised  by non-coherent
periods dominated by the fundamental and first harmonic  with variable relative
amplitudes.  By comparing the equatorial  rotational velocities obtained from 
the photometric fundamental periods  with the projected rotational velocities, 
it is shown that the fundamental  period is indistinguishable from the rotation
period.  The distributions of the projected rotational velocities and the 
equatorial rotational velocities for Be and non-Be stars are discussed.  Using 
the equatorial rotational  velocities of Be stars directly determined from the 
photometric period, it is  found that, on average, Be stars rotate at about 
0.66 of critical velocity. This result is independent of assumptions concerning
gravity darkening.  It  is shown that the rotational amplitudes of Be stars 
increase with effective  temperature and are considerably higher than those of 
non-Be stars.  The  amplitude distributions of Be and non-Be stars are also 
significantly different.   Of particular interest is a large rise in rotational
amplitudes from A0 to B5  for non-Be stars.  We suspect that this may be a 
result of the increasing  importance of a line-driven wind.  Perhaps the Be 
stars are just those  with greatest surface activity with increased mass loss 
aided by rapid rotation. 
\end{abstract}


\section{Introduction}

One of the most important unsolved problems in astrophysics is the mechanism 
which leads to episodic mass loss in Be stars. \citet{Struve1931} was the first
to suggest that Be stars rotate at critical velocity, forming lens-shaped 
bodies which eject matter at the equator. However, a statistical analysis of 
the distribution of projected rotational velocities, $v\sin i$, showed that 
the equatorial rotational velocities of Be stars are significantly below 
critical \citep{Slettebak1949}.  \citet{Stoeckley1968} suggested that
$v\sin i$ measurements were perhaps too low because the effects of 
gravitational darkening and spherical distortion might have been 
under-estimated.  He showed that the $v\sin i$ distribution of 40 Be stars 
could be consistent with critical rotational velocity.  An analysis of a
larger sample of Be stars by \citet{Slettebak1979} offered no support for
critical rotational velocity, though it could not be entirely excluded.

It has long been known that Be stars vary photometrically and spectroscopically
over long time scales which can be attributed to the circumstellar disk.  The 
first report of short-term periodic variability in a Be star was by 
\citet{Walker1953} who found a 0.73--0.80\,d light variation in EW~Lac.  
\citet{Baade1979, Baade1982} found spectroscopic line profile variations with 
a 1.36-d period in the Be star 28~CMa.  This was interpreted as nonradial 
pulsation (NRP).  From that time onwards, the idea of NRP as a trigger for 
mass ejection began to gain increasing acceptance.

NRP in B stars is attributed to the opacity mechanism acting in the partial
ionization zone of iron-like elements and is responsible for coherent
pulsations in $\beta$~Cep and SPB stars.  The amount of momentum transfer that 
can be applied by NRP is limited by the sound speed in the stellar atmosphere. 
For that reason, a star needs to be rotating in excess of 90 percent of the 
critical rotation velocity for material to be ejected \citep{Townsend2004}.  In
order to justify near-critical rotation,  Stoeckley's idea concerning the 
extreme effect of gravitational darkening at near-critical velocity 
was resurrected \citep{Townsend2004}.  

In this hypothesis, all Be stars are rotating near critical velocity and must, 
at times, pulsate with sufficient amplitude to drive the mass loss.  It was 
originally suggested that beating of two or more pulsation modes would generate 
the required amplitude.  However, the evidence for the predicted periodic 
outbursts does not exist except perhaps in the case of $\mu$~Cen 
\citep{Rivinius1998b,Baade2016}.  Beating of coherent modes would in any
case be expected to lead to a rather slow increase in amplitude, while Be 
outbursts, as its name implies, are impulsive events. 

Recent analyses of the  $v\sin i$ distribution have consistently found no 
evidence that Be stars are rotating close to critical velocity, $v_c$.  
\citet{Yudin2001} finds that early-type Be stars rotate at typically
0.5--0.7$v_c$, while late-type Be stars at about 0.7--0.8$v_c$.
\citet{Cranmer2005} found that early-type Be stars rotate at about 
0.4--0.6$v_c$, though a few may be rotating near critical velocity.  Late-type 
Be stars exhibit progressively narrower ranges of rotation speed with 
decreasing effective temperature with the coolest Be stars rotating close to 
critical. \citet{Zorec2016} has arrived at the same conclusion, showing that 
Be stars generally rotate at about $0.6v_c$, but with a range 0.3--0.95\,$v_c$.
From spectro-interferometric observations of 26 Be stars, \citet{Cochetti2019}
find that they rotate at about $0.75 v_c$. 

\citet{Walker1953} suggested that the short-period variation in EW~Lac may
be due to ``temporary disturbances which form on or near the surface of the star 
and are carried across its disk by the stellar rotation. The passage of a dark, 
cooler region across the visible hemisphere of the star could account for the 
light-variation and possible color variations, as well as for the change in the
amplitudes of the minima, as the disturbance grew and diminished in size
and intensity''.  Essentially, this can be thought of as a co-rotating cloud of
gas which eventually disperses into the circumstellar disk.  The circumstellar
disk material certainly has a large effect on the light and line profile
variations, so this conclusion seems reasonable.  After all, any
material ejected by the star (including by NRP) cannot avoid occulting the 
photosphere.  If the material lingers near the photosphere for a few days in a 
Kepler-like orbit or trapped by a magnetic field this will inevitably lead to 
periodic light variations. 

The reason why Walker's idea was rejected and replaced by NRP is that stars 
with radiative envelopes have been assumed to have immaculate, inactive, 
photospheres.  It is generally believed that magnetic fields, which are
responsible for activity in the Sun and other cool stars cannot be generated
in radiative envelopes. This long-held view can now be questioned owing to
recent photometric {\it Kepler} and {\it TESS} observations of A and B stars.  
Reports of flares in A and late-B stars \citep{Balona2012c,Balona2015a} as well
as rotational modulation in a large fraction of A and B stars 
\citep{Balona2013c, Balona2016a, Balona2017a, Balona2019c} suggest that 
activity is indeed present.  Furthermore, new ideas regarding 
how surface magnetism and starspots may be formed in early type stars have 
been recently proposed \citep{Cantiello2009, Cantiello2011,Cantiello2019}. 
This opens the way for a new interpretation of the short-period variations
and mass loss in Be stars.

\begin{figure*}
\begin{center}
\includegraphics{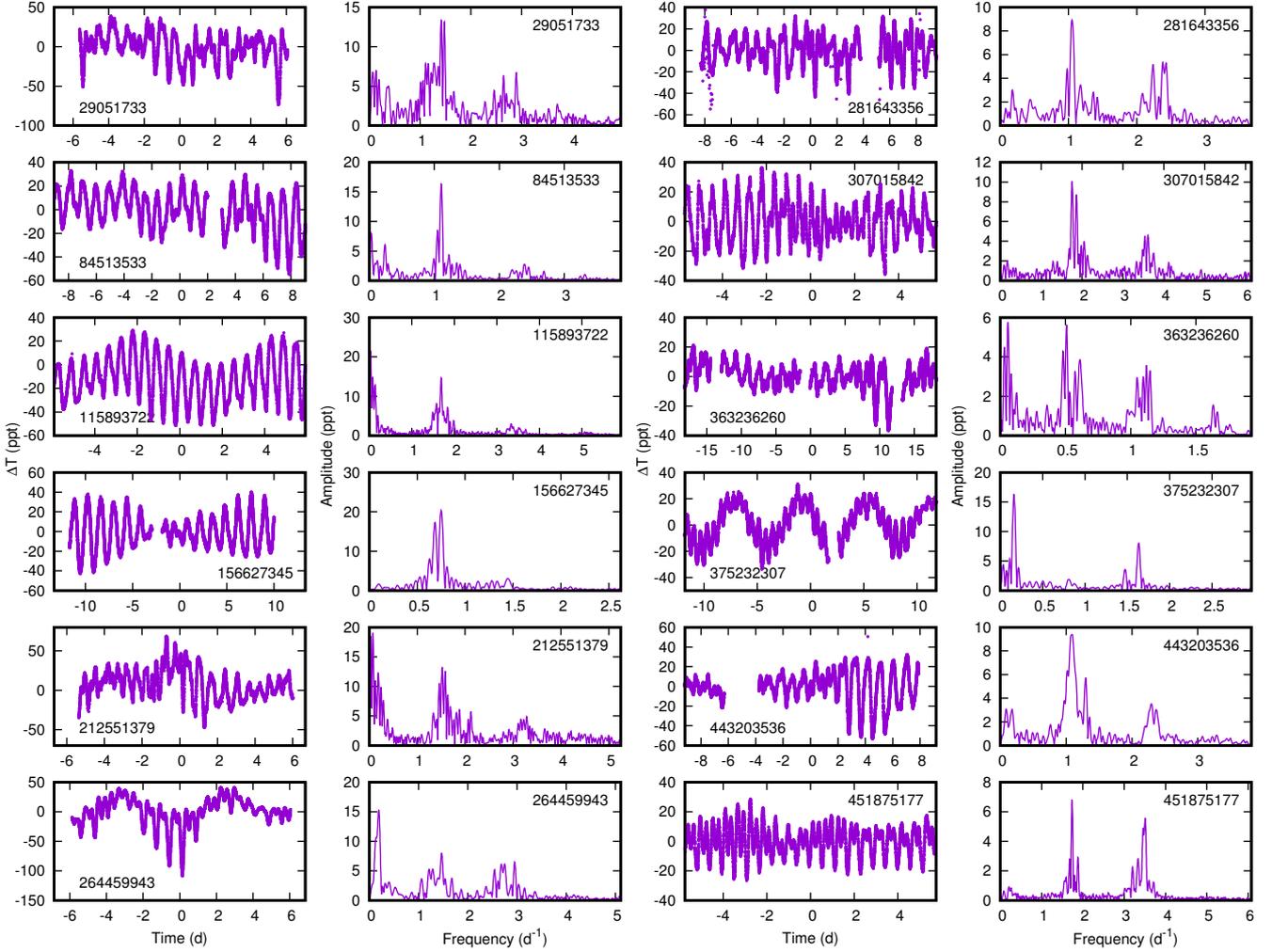}
\caption{Parts of the light curves and periodograms of the full light curves
of some Be stars.  The brightness and amplitudes are in parts per thousand 
(ppt).  The TIC numbers are shown.}
\label{belc}
\end{center}
\end{figure*}

\citet{Balona1990, Balona1995} found that the relationship between $v\sin i$ 
and the photometric period in Be stars is consistent with rotational modulation.  
\citet{Balona2003b} proposed a magnetic rotator model for mass loss which 
involves the expulsion of material from an active region in the photosphere.
The expelled material is directed to two diametrically opposite regions where 
the geometric and magnetic axes intersect and forced into co-rotation with the 
star.  This eventually leads to a build-up of material at the two locations, 
leading to a single- or double-wave light curve depending on the relative 
amounts of trapped gas at the two regions. Eventually, the gas disperses into 
the circumstellar disk.  We will call this the ``impulsive magnetic
rotator'' model.

In large part, the model was driven by observations which clearly show that at 
times Be stars can change from a single wave to a double wave light curve 
\citep{Balona1986c, Balona1991b,Balona2020b}.  This has been quite a well-known
characteristic of Be stars for many decades. It is often expressed as frequency
groupings in the periodogram, with one group at double the frequency of the 
other.  The relative amplitudes of the two groups vary from season to season.
The model was also driven by the fact that early-type Be stars are more
active than late-type Be stars.  In the model this is partly due to the 
increasing importance of line-driven stellar winds with increasing effective 
temperature.

In our first paper \citep{Balona2020b}, we discussed {\em TESS} observations
of individual Be stars of particular interest.  In this paper we examine all 
{\em TESS} classical Be stars in quarters 1--26.  The aim is to investigate 
the properties of the light curves and to test the model proposed by 
\citet{Balona2003b}.  As part of a project to classify all {\em TESS} stars 
hotter than 6000\,K according to variability type, the fraction of Be stars 
showing short-period variations is determined.  Stellar radii are estimated 
from {\it Gaia}\,DR2 parallaxes and effective temperatures.  Together with the 
photometric period, the presumed equatorial rotational velocity, $v$, can be 
found.  Values of $v$ are compared with $v\sin i$ in a test for rotational 
modulation.  The $v$ distribution is compared with the $v\sin i$ distribution 
as a further test. The distribution of $v$ and the distribution of $v/v_c$ for 
Be stars is obtained without de-convolution for the first time.  Finally, the
rotational light amplitudes of Be stars are compared with those of non-Be 
stars at the same effective temperature to determine if there is a difference 
in how the variations are produced in Be and non-Be stars.

\section{The data and catalogue}

{\em TESS} obtains continuous wide-band photometry with two-minute cadence
over 13 observing segments per hemisphere.  Each segment is observed
continuously for 27\,days.  Because sectors overlap, stars near the ecliptic
equator are only observed for one sector.  Stars in the circular regions where 
segments overlap at the ecliptic poles are observed for about 100\,days. 
The light curves are corrected for time-correlated instrumental signatures
using pre-search data conditioning (PDC, \citealt{Jenkins2016}).  The data
used here are from sectors 1--26.

This project forms part of a larger project to create a catalogue of the
variability type  for every star hotter than 6000\,K observed by {\em TESS} 
and {\em Kepler}. The variability classification follows that of the 
{\em General Catalogue of Variable Stars} (GCVS, \citealt{Samus2017}).  The 
only recognized class of rotational variable among the A and B stars are the 
chemically peculiar $\alpha^2$~CVn and SX~Ari classes.  A new ROT class has been 
added to describe any star in which the variability is suspected to be due to 
rotation and not known to be Ap or Bp.  The classification was accomplished by 
visual inspection of the light curve and Lomb-Scargle periodogram 
\citep{Scargle1982} and a rough estimate as to whether the star is hotter or 
cooler than type A0.  This is necessary in order to distinguish $\beta$~Cep 
and Maia stars from $\delta$~Sct variables and SPB from $\gamma$~Dor stars.  
Aided by suitable software, classification of over 100 stars an hour is 
possible.  

For each star, the literature was searched for the best estimate of the 
effective temperature, $T_{\rm eff}$, projected rotational velocity, $v\sin i$,
and spectral type.   Values of $T_{\rm eff}$ in the {\em Kepler Input 
Catalogue} (KIC, \citealt{Brown2011a}) and the {\em TESS Input Catalogue} 
(TIC, \citealt{Stassun2018}) are unreliable for B stars because the photometric
observations from which they are derived lack the U band.  Whenever
possible, estimates from spectroscopic modelling are used.  Values of $T_{\rm
eff}$ for Be stars are based on the spectral type because even spectroscopic
modelling is unreliable due to emission.  For this purpose, the 
\citet{Pecaut2013} calibration is used.  The resulting uncertainty in 
$T_{\rm eff}$ is probably 1000--2000\,K.

The stellar luminosity was estimated from {\it Gaia} DR2 parallaxes  
\citep{Gaia2016, Gaia2018} in conjunction with reddening obtained from a
three-dimensional map by \citet{Gontcharov2017} using the bolometric correction 
calibration by \citet{Pecaut2013}. From the error in the {\it Gaia} DR2 
parallax, the typical standard deviation in $\log(L/L_\odot)$ is estimated to 
be about 0.05\,dex, allowing for standard deviations of 0.01\,mag in the 
apparent magnitude, 0.10\,mag in visual extinction and 0.02\,mag in the 
bolometric correction in addition to the parallax error.  Over 61000 {\em TESS}
stars and 22000 {\em Kepler} stars have been classified.

\section{Short-period variability in Be stars}

The Be stars in this paper are identified using the {\em BeSS} 
database \citep{Neiner2011}.  This is a catalog of 2330 classical Be stars.
Of the 441 classical Be stars in sectors 1--26 observed by {\em TESS}, 319 
stars (i.e. 73\,percent) were classified as ROT variables.  Clearly, 
quasi-periodic low frequencies are a general characteristic of most Be stars.  

It is relatively easy to identify Be stars from the light curve and
periodogram.  The general appearance of the light curve tends to be a 
mixture of fundamental and first harmonic with variable amplitudes.  An 
additional distinction is that many Be stars show a long-term drift in 
brightness which is sometimes sudden and erratic and, in a few cases, 
quasi-periodic (e.g., TIC\, 375232307 in Fig.\,\ref{belc}). This is not seen in 
non-Be stars.  About half of the 441 Be stars can be identified from the
light curve alone.

\begin{figure}
\begin{center}
\includegraphics{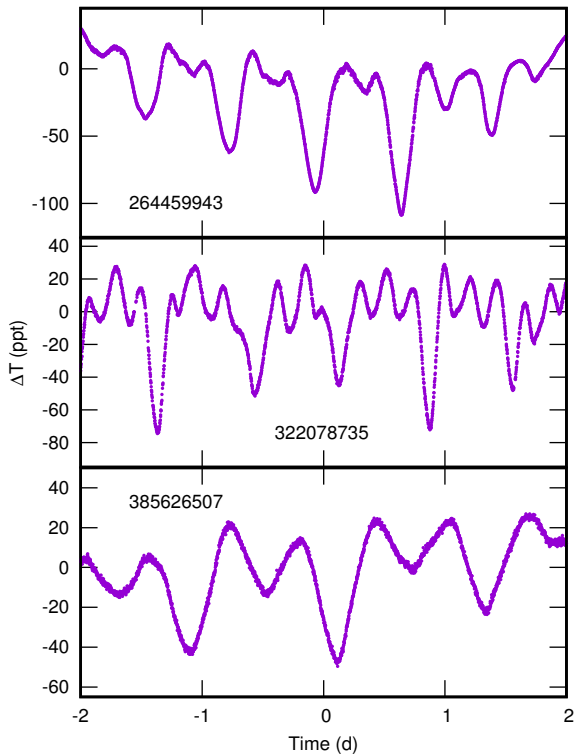}
\caption{Detail of the light curves showing periodic eclipse-like dimming
events.}
\label{eclipse}
\end{center}
\end{figure}

Fig.\,\ref{belc} shows examples of Be light curves and corresponding 
periodograms.  Note that the fundamental and first harmonic are nearly always 
present.  The frequency peaks are broad, indicating non-coherent variations.  
Another identifying feature in the light curves of many Be stars is that the 
quasi-periodic variations sometimes resemble eclipses.  In other words, the 
light curve is not sinusoidal, but consists of regular light dips from a more
slowly varying light maximum.  Examples are shown in Fig.\,\ref{eclipse}.

Although the light curves and periodograms of most Be stars fit the above
description, 92 stars in the {\it BeSS} catalogue do not. Some are eclipsing 
binaries, while others are $\beta$~Cep, SPB, Maia or $\delta$~Sct pulsating 
variables.  In others, where no short-period variations are seen, the 
irregular light variations can probably be attributed to the circumstellar disk.

Groupings of fundamental and first harmonic seem to be present in some B stars 
not known to be Be stars \citep{Balona2011b}.  Indeed, 11 {\em TESS} stars not 
known to be Be, have light curves and periodograms closely resembling those of 
Be stars.  Of these, 3 are Bn stars.

\section{Rotational modulation}

As mentioned in the Introduction, a large fraction of normal A and B stars
appear to show low-amplitude periodic light variations which are difficult
to understand unless this is a result of starspots \citep{Balona2019c}. 
However, the light variations in Be stars seem to be far more erratic and of 
larger amplitude than those seen in A and late B stars.  The frequency peaks
in Be stars are sometimes very broad (Fig\,\ref{belc}), whereas those in A 
and late B stars are generally narrow.  The rapid changes in amplitude
of the relative strengths of fundamental and first harmonic in Be stars, as
discussed above, are never seen in non-Be stars.

These differences indicate that the cause of rotational modulation in non-Be
stars is different from that in Be stars.  The observations suggest that
whereas the features are rather stable and long-lived in non-Be stars, this
is not the case in Be stars.  In the impulsive magnetic rotator model, the
non-coherence of the light variations can be understood because the trapped
gas clouds are continuously changing shape and size.  In this model it is the 
fundamental period which represents the  rotation period.

One can test if the fundamental period is the same as the rotation period
by comparing the projected rotational velocity, $v\sin i$, with the equatorial 
rotational velocity, $v$, estimated from the stellar radius.  The stellar 
radius can be obtained if we know $T_{\rm eff}$ and the luminosity.

\begin{figure}
\begin{center}
\includegraphics{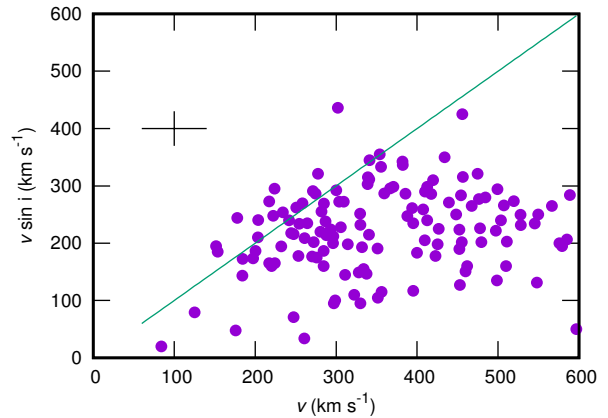}
\caption{The projected rotational velocity, $v\sin i$, as a function of the
equatorial rotational velocity, $v$, for 132 Be stars observed by {\it
TESS}.  The straight line represents $\sin i = 1$.  Typical error bars for
each point are shown by the cross.}
\label{vsini}
\end{center}
\end{figure}

In the first test, shown in Fig.\,\ref{vsini}, 132 Be stars with known
$v\sin i$ and with {\it Gaia} luminosities are shown.  The equatorial radius
in a rapidly-rotating star is larger than the mean radius.  Therefore the
estimated equatorial rotational velocity, $v$, is probably too low.  On the 
other hand, the value of $T_{\rm eff}$ derived from the spectral type reflects
mostly the cool equatorial region, which means that the resulting $T_{\rm eff}$ 
is perhaps somewhat lower than the true $T_{\rm eff}$, resulting in a
somewhat larger value of $v$.  These two effects balance each other. 
Considering the large uncertainty in the stellar radius, it is probably best 
to leave $v$ unchanged.   

If the photometric period in Be stars is the rotation period, the measured
$v\sin i$ should be lower or equal to the equatorial rotational velocity,
$v$, though some stars may be expected to lie above the $\sin i = 1$ curve
due to uncertainties in the stellar parameters.  Also, most stars should be 
near the $\sin i = 1$ line because more stars are likely to be seen equator-on 
rather than pole-on.  Fig.\,\ref{vsini} does indeed show the expected behaviour.  
We conclude that the photometric periods are indistinguishable from the 
rotation periods.

\begin{figure}
\begin{center}
\includegraphics{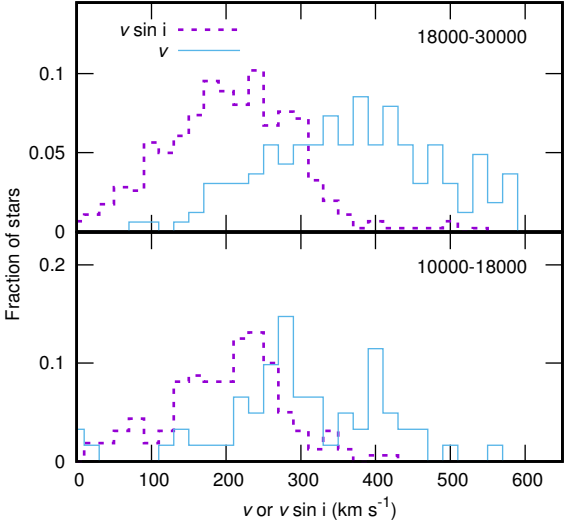}
\caption{The distribution of equatorial rotational velocities of Be stars
(blue) and the distribution of projected rotational velocities, $v\sin i$
for two effective temperature ranges.}
\label{prot}
\end{center}
\end{figure}

Another test can be made by comparing the distribution of $v\sin i$ with the
distribution of $v$.  For this test, it is reasonable to use all available
measurements of $v\sin i$ for Be stars in the literature.  In this way 
$v \sin i$ measurements of 461 classical Be stars with $18000 < T_{\rm eff} 
< 30000$\,K and 160 classical Be stars with $10000 < T_{\rm eff} < 18000$\,K
can be used (Fig.\,\ref{prot}).

In principle, one could deconvolve the $v\sin i$ distribution to obtain the 
distribution of equatorial rotational velocities, $v$, assuming random
orientations of the axes of rotation.  However, the orientation of the
rotation axes of stars used to obtain the $v$ distribution cannot be random. 
The amplitude of rotational modulation will decrease with decreasing 
inclination and will vanish when the star is pole-on, even though $v\sin i$ can
still be measured.  For this reason, the deconvolved $v$ from $v\sin i$ and the
photometric $v$ distributions cannot be the same.

In Fig.\,\ref{prot}, values of $v$ from 164 stars were used for the range
$18000 < T_{\rm eff} < 30000$\,K and 61 stars for the range $10000 < 
T_{\rm eff} < 18000$\,K.  Considering the large error in the estimated $v$ and 
the relatively small numbers of stars, not much weight can be placed on 
the details.  The $v$ distribution is clearly displaced to higher velocities, 
as expected.  \citet{Chandrasekhar1950} derived relations between the moments 
of the $v$ and $v\sin i$ distributions assuming random orientation of the
rotation axes.  For the mean, they find $\langle v\rangle = 1.27\langle 
v\sin i\rangle$. For 93 stars where both measurements are available, we obtain 
$\langle v\sin i\rangle = 229 \pm 8$\,km\,s$^{-1}$ and $\langle v \rangle = 
370\pm 12$\,km\,s$^{-1}$ for $18000 < T_{\rm eff} < 30000$\,K.  Thus 
$\langle v\rangle = 1.62\langle v\sin i\rangle$.  The mean $v$ determined
from photometry will be larger than the mean $v$ derived from deconvolution of 
$v\sin i$ because there will be a deficit of stars with low rotation rates 
among the photometric $v$ values.  Thus the larger value of the proportionality 
factor is to be expected.

\begin{figure}
\begin{center}
\includegraphics{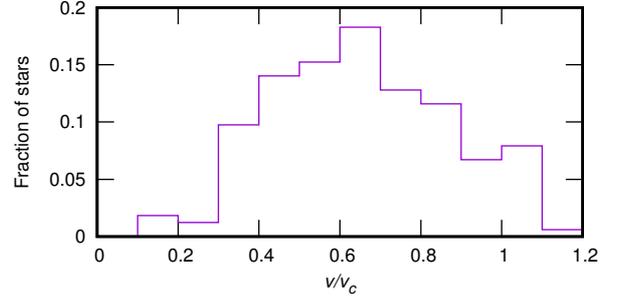}
\caption{The distribution of $v/v_c$ for Be stars with $18000 < T_{\rm eff} <
30000$\,K.}
\label{vevc}
\end{center}
\end{figure}

From the photometric $v$ distribution it is possible to obtain an idea of
the ratio of equatorial velocity to critical velocity, $v/v_c$. Using
$T_{\rm eff}$, $\log\,g$ (estimated from the luminosity), and the
metallicity (assumed solar), the stellar mass may be derived using 
the relation in \citet{Torres2010a}.  From the mass and radius, $v_c$ is
obtained.  The distribution of $v/v_c$ for early-Be stars is shown in
Fig.\,\ref{vevc}.  The mean value from 225 stars is $\langle v/v_c\rangle = 
0.66 \pm 0.02$. This value is the same as that derived by \citet{Zorec2016} 
from deconvolution of $v\sin i$.  Note that there is no gravity darkening
dependence and the result is independent of $v \sin i$.  This means that it 
can no longer be argued that the $v\sin i$ values are too low owing to
extreme gravity darkening in Be stars \citep{Townsend2004}.

\begin{figure}
\begin{center}
\includegraphics{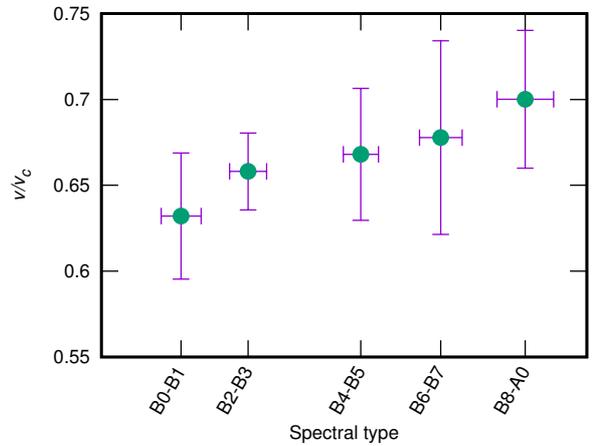}
\caption{The ratio of equatorial rotational velocity to critical rotational
velocity, $v/v_c$, as a function of spectral type for Be stars.
The bars are 1-$\sigma$ standard deviations.}
\label{vrat}
\end{center}
\end{figure}

Fig.\,\ref{vrat} shows that $v/v_c$ tends to increase slightly towards late Be 
stars.  This result has been known for a long time \citep{Slettebak1982,
Yudin2001, Cranmer2005}. \citet{Cranmer2005} found $v/v_c$ increases from 
0.6 to 0.7 between early- and late-Be stars, similar to what is found here.

\begin{figure}
\begin{center}
\includegraphics{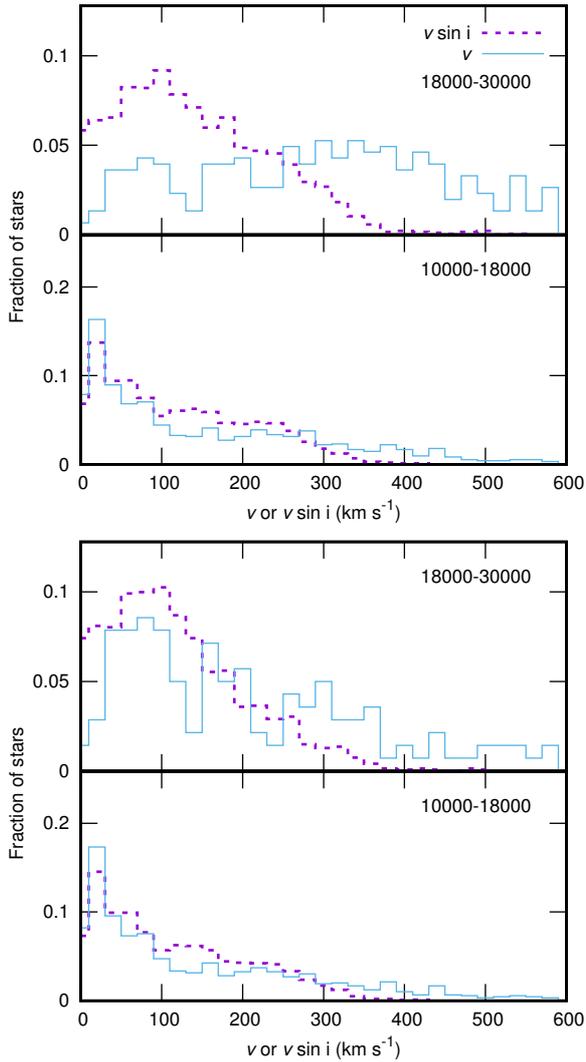}
\caption{The top panels show the distribution of equatorial rotational 
velocities (solid blue line) and of $v\sin i$ (dotted violet line) for all 
main sequence B stars, including Be stars, in the indicated range of
effective temperatures.  The bottom panels show the same distributions,
excluding Be stars.}
\label{dist}
\end{center}
\end{figure}

\begin{figure}
\begin{center}
\includegraphics{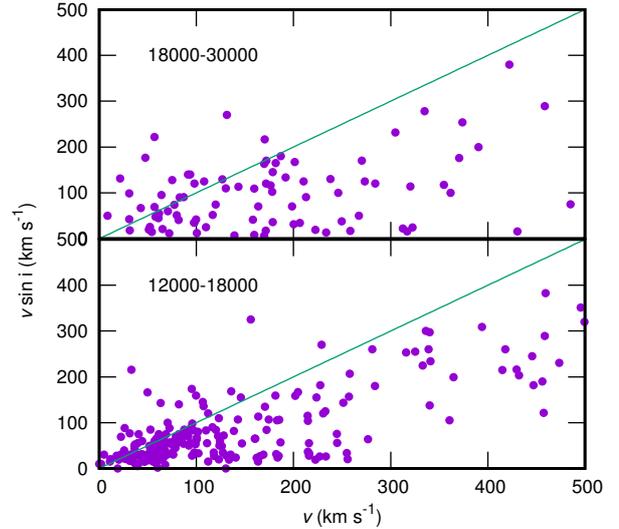}
\caption{The projected rotational velocity, $v\sin i$ as a function of the
equatorial rotational velocity, $v$, for non-Be main sequence B stars
in two effective temperature ranges. The straight lines have unit 
slope.}
\label{vnon}
\end{center}
\end{figure}

It is instructive to compare the photometric and spectroscopic rotational 
velocity distributions when non-Be stars are included.  Because Be stars
are rapid rotators, an excess of Be stars with known $v$ relative to the 
general population will distort the distribution.  The top panels of 
Fig.\,\ref{dist} shows the $v$ and $v\sin i$ distributions for all B stars, 
including Be stars.  For the 18000--30000\,K range, the $v$ distribution is 
derived from 304 Be stars and 140 non-Be stars.  In other words, 68\,percent 
are Be stars.  The $v\sin i$ distribution for the early B stars is derived 
from 1939 Be stars and 1482 non-Be stars.  This means that 57\,percent of 
stars defining the $v\sin i$ distribution are Be stars.  This difference may 
account for the poor agreement.

To test this idea, one may compare the $v$ and $v\sin i$ distributions
excluding Be stars in both cases.  As shown in the bottom panels of 
Fig.\,\ref{dist},  the agreement between the two distributions is improved.
For the early-B stars, the $v$ distribution is obtained from 140 stars and 
the $v\sin i$ distribution from 1482 stars.  This seems to support the idea 
that it is the inclusion of Be stars that distorts the $v$ distribution for
early-B stars. The distributions for the late-B stars seem unaffected by the 
inclusion or exclusion of Be stars.

To strengthen the idea that rotational modulation is present among a large 
fraction on Be and non-Be stars, we show in Fig.\,\ref{vnon} the
relationship between $v\sin i$ and $v$ for main sequence non-Be stars in two
temperature ranges.  It is clear from Fig.\,\ref{vnon}, that the photometric 
periods are the same as the rotation periods, suggesting the presence of 
starspots or similar obscuration in non-Be main-sequence stars.

A more accurate measure of the relative proportion of Be stars among B stars
may be obtained by counting all main sequence B stars brighter than 10-th 
magnitude.  This limit was chosen on the assumption that the sample of Be 
stars is probably close to completion.  Within this brightness range, there are 
1183 stars with $18000 < T_{\rm eff} < 30000$\,K of which 282 are Be stars.  
Thus Be stars comprise about 24\,percent of all early-B stars.  As mentioned 
above, the ratio of Be stars which were classified as ROT is 68\,percent. This 
large fraction of ROT stars in the early-B range clearly suggests that it is 
much easier to detect rotational modulation among Be stars than in non-Be 
stars.

Of the 1534 stars with $10000 < T_{\rm eff} < 12000$\,K brighter than 10-th 
mag, 92 are Be stars (6\,percent).  This explains why the velocity
distributions for late-B stars are relatively unaffected by the inclusion or 
exclusion of Be stars.

The comparison of the $v$ and $v\sin i$ distributions for early-B stars
seems to suggest that the light variations in Be stars are of a different 
nature from those in non-Be stars.  Rotational modulation in Be stars is
much easier to detect than in non-Be stars, presumably because they have
larger amplitudes (see below).  This lends support to the idea that rotational 
modulation in non-Be stars is a result of starspots, whereas in Be stars it is 
a result of co-rotating clouds.  Additional support may be found in the very 
important {\it CoRoT} time series of TIC\,234230792 (HD\,49330) first reported 
by \citet{Huat2009} and described by \citet{Balona2020b}.  This star happens to
be a $\beta$~Cep pulsating variable as well as a classical Be star.  There
is evidence that the reduction and subsequent increase in amplitude of the
pulsations is a consequence of obscuration by co-rotating clouds resulting
from an outburst.

\begin{figure}
\begin{center}
\includegraphics{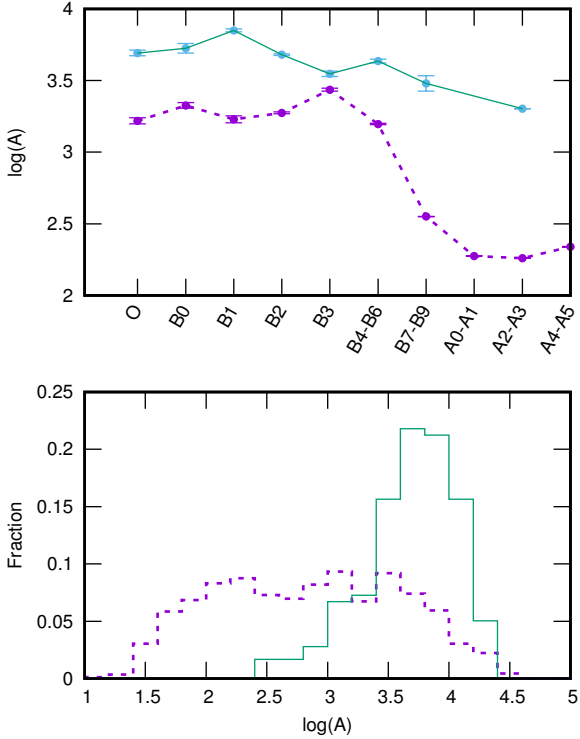}
\caption{Top panel: mean amplitudes (in ppm) of Be stars (green solid line) 
and non-Be stars (violet dotted line) as a function of spectral type.  The
1-$\sigma$ standard errors are shown.  Bottom panel: the distribution of 
amplitude (ppm) of B0--B9 stars for Be stars (solid green histogram) and 
non-Be stars (violet dotted histogram).}
\label{ampdis}
\end{center}
\end{figure}

\section{Amplitudes}

While the quasi-periodic light and line profile variation in early-Be stars is
detectable from ground-based observations, the periodic variations
associated with starspots in non-Be stars are only detectable from
space.  Apparently, the obscuration which causes the short-period
variations in Be stars is much larger than the obscuration in non-Be stars.
To quantify the comparison, the amplitudes of the fundamental and first 
harmonic in {\it TESS} Be stars were added in quadrature and taken as 
representative of the amplitudes of the quasi-periodic variation.  The same 
was done for non-Be stars and the two compared as a function of spectral type 
(top panel of Fig.\,\ref{ampdis}).  

The amplitudes of the Be stars decrease towards later types, which has been
known for a long time.  Line-profile variations among late-Be stars are very
difficult to detect \citep{Baade1989}.  It is only with the advent of space 
photometry that short-period light variations were finally detected in late Be 
stars.  Even though the photometric amplitudes of late-Be stars are small,
they are still about an order of magnitude larger than non-Be stars.
What is particularly noteworthy is the rapid increase in rotational amplitude 
between A0 and B5 for non-Be stars (top panel of Fig.\,\ref{ampdis}). 
This clearly requires some thought.

In the bottom panel of Fig.\,\ref{ampdis}, the amplitude distribution for
Be stars is compared with non-Be stars.  To facilitate the comparison,
this is plotted as a log-normal distribution.  What is clearly evident is
that the amplitude distribution for Be stars is strongly peaked and quite 
different from that of non-Be stars. 

It is possible that the large rotational amplitudes for non-Be stars earlier
than B5 may partly be a result of undetected Be stars in the sample.  One idea
is that there is a continuum between non-Be and Be stars.  Perhaps there is
a range of surface activity and it is only the most active and
rapidly-rotating stars in this range that appear as Be stars.  It is well-known
that a Be star can lose its emission for many years and then regain it again.  
This is very difficult to understand unless there is a mechanism which allows 
for variation in the mass loss.  Variations in surface activity may be such a 
mechanism.

\section{Discussion and Conclusions}

In \citet{Balona2020b} {\em TESS} light curves of 57 classical Be stars were
discussed on a star-by-star basis.  It was found that most Be stars have 
non-coherent short-period variations in which the fundamental and first 
harmonic are dominant.  The fundamental period was shown to be most likely the 
same as the rotation period.  The impulsive magnetic rotator model proposed by 
\citet{Balona2003b} was used as a framework for understanding the general 
characteristics of Be stars.

In this paper, we examine a much larger sample of 463 classical Be stars
from {\em TESS} sectors 1--26 with the aim of extracting the most important
characteristics of the class as a whole.  

It is found that about 73\,percent of Be stars in this sample
have short-period light variations. Many Be stars can be identified from their 
characteristic light curves and periodograms.  In the periodograms of most Be 
stars, the fundamental and first harmonic peak are often visible.  The peaks
are always broadened to greater or lesser extent, indicating a non-coherent
quasi-periodic behaviour.  This is in contrast to the sharp peaks expected
from self-driven pulsations.  The relative amplitudes of the fundamental and 
first harmonic may vary on quite short timescales, appearing as a double-wave 
light curve with components of variable amplitude.  In some cases the 
fundamental or first harmonic disappears altogether, only to reappear at a 
later time \citep{Balona1991b,Balona1986c, Balona2020b}.  

From the quasi-period and a radius estimate using {\em Gaia}\,DR2 parallaxes, 
the equatorial rotational velocity, $v$, can be found.  The correlation between the equatorial 
$v$ and the projected rotational velocity, $v\sin i$, for 132 Be stars confirms 
that the fundamental quasi-period is indistinguishable from the rotation 
period.  The distribution of equatorial rotational velocities of early-type Be 
stars is directly derived.  It is found that Be stars typically rotate at 
about 0.6--0.7 of the critical velocity, but with a wide range of velocities.  
The same result has repeatedly been obtained by several previous analyses
\citep{Yudin2001, Cranmer2005, Zorec2016, Cochetti2019}.  

An analysis of the ratio of equatorial rotation velocity to critical
rotational velocity, $v/v_c$, shows that $\langle  v/v_c\rangle = 0.66$ with
a wide spread in $v/v_c$.  This result is independent of any assumption
regarding gravity darkening and shows that all Be stars cannot possibly be
rotating close to critical.

From space photometry of A and B stars, it has been shown that about
20--40\,percent of hot main sequence stars show rotational modulation,
presumably indicating the presence of starspots \citep{Balona2013c,
Balona2019c, Balona2020p}.  However, the rotational light amplitudes of Be 
stars seem to be considerably larger.  The amplitude distributions of Be and
non-Be stars also differ.  It is proposed that in Be stars the periodic
light and line-profile modulation is due to magnetically trapped gas clouds
rather than starspots.  This would also explain the greater non-coherence of 
the short-period variations in Be stars.

The early-type Be stars have larger rotational amplitudes than late Be stars.  
Furthermore, even among non-Be stars, there is a sharp rise in rotational
amplitudes for A0-B5 stars.  This may perhaps be attributed to the larger role 
played by line-driven winds in the early-B stars. 

It is speculated that the difference between non-Be and Be stars lies in the
amount of surface activity and the rotation rate.  Only the most active and
rapidly-rotating stars attain a mass loss rate which is sufficient to cause
measurable H$\alpha$ emission.  A loss in activity would lead to a reduced
mass loss rate and the loss of emission, thus explaining why emission may
disappear and reappear in a Be star. 

The observations may be understood in terms of the impulsive magnetic
rotator model \citep{Balona2003b}.  This model is similar to the centrifugal 
magnetosphere model \citep{Babel1997, ud-Doula2002}.  However, most of the
mass loss is not due to a line-driven wind, but to an expulsion of material 
from active regions associated with starspots. This accounts for the impulsive 
nature of Be outbursts.  Furthermore, the tilted magnetic field is very weak 
(perhaps only a few Gauss).   The expelled material is channeled to two 
diametrically opposed locations where the geometric and magnetic equators 
intersect.  Material located in these two co-rotating regions leads to 
non-coherent rotational modulation (quasi-periodicity) owing to short-time 
changes in size and density of the trapped clouds.  

The presence of two diametrically located regions of obscuration explains
why the first harmonic is such a characteristic feature of Be stars.  Changes 
in the relative amount of material trapped in the two regions will account for 
the large changes in relative amplitude of fundamental and first harmonic in 
Be stars.

An early-B star will have a large ionization radius.  Even if the star is 
rotating well below critical velocity, the long lever arm provided by a weak 
magnetic field will ensure that the ejected matter attains circular velocity 
while still ionized and under control of the magnetic field.  A late Be star
will have a much shorter ionization radius, which means that material
remains ionized and under control of the magnetic field much closer to the 
photosphere. The lever arm is therefore shorter and requires a star already 
rotating closer to critical for the gas to reach circular velocity.
This might explain why late Be stars are rotating in a narrow band close to 
the critical rotation speed \citep{Yudin2001, Cranmer2005, Zorec2016}.

The condition to attain co-rotation velocity is much easier to attain in
early-B stars than in late B stars because of the larger ionization radius
and the importance of line-driven winds.  This might explain why most Be stars 
are of early type.  In this way, one can also understand why early-Be stars 
are more numerous and more active than late-Be stars.

While rotational modulation was proposed long ago \citep{Walker1953,
Balona1990, Balona1995}, it was not generally accepted because it was assumed 
that stars with radiative envelopes have immaculate, tranquil outer layers 
which cannot provide the dynamo action required to generate a surface magnetic 
field.  As a result, non-radial pulsation became the most widely accepted
model for triggering mass loss \citep{Baade1979, Baade1982}.  While it does
explain the short-period variations (though the implicit assumption of
frequency coherence is not supported by observations), the prediction of 
periodic episodic events of mass loss due to beating of modes has failed the 
test \citep{Rivinius1998b,Baade2016}. The assumption of near-critical 
rotational velocity for all Be stars has also failed the test 
\citep{Yudin2001, Cranmer2005, Zorec2016, Cochetti2019}. 

It is true that in some Be stars coherent light variations are to be found.
The above model by no means rules out that some of the photometric and line 
profile variations in Be stars are indeed a result of NRP.  There are several 
instances of Be stars which are $\beta$~Cep variables and there are, no doubt, 
Be stars which are SPB variables as well.  After all, they do occupy the
$\beta$~Cep and SPB instability strips and there is no reason why pulsation
should not occur in an impulsive magnetic rotator.  However, NRP requires 
coherent pulsations and cannot explain the incoherent variations
characteristic of Be stars.

It is suspected that Be stars may be just one extreme of a continuum
depending on the strength of the field.  While Be stars are suspected to
have very weak fields, the magnetic Bp stars, which have very strong tilted 
dipole magnetic fields, resulting in surface element segregation, are the 
other extreme. In the hottest of these stars, mass loss is dominated by 
line-driven winds.  This model has been successfully used by 
\citet{Townsend2005b} for prediction of the light curve of the He-rich 
chemically peculiar star $\sigma$~Ori~E. 

It is often claimed that a magnetic field has not been detected in any Be
star, ruling out the impulsive magnetic rotator model.  This is a disingenuous 
argument because a field of the order of one Gauss is all that is required
in the model.  At present, the most accurate measurements on Be stars have 
errors measured in tens of Gauss \citep{Wade2016b}..  In spite of this, 
HD~208057 (16~Peg) a classical Be star with a rotation period of 1.37\,d, has 
a field strength $B_z = 210\pm 50$\,G \citep{Wade2016}. \citet{Neiner2012b} 
investigated the Be star $\omega$~Cen and obtained an upper limit of about 
80\,G (the median error of the measurements is 30\,G).  \citet{Neiner2012b}
also found evidence for co-rotating clouds.  It is imperative that
measurement errors be lowered significantly before any claims of the lack of
magnetic fields in Be stars can be made.

Unless a different model is proposed for the presence of rotational modulation 
in A and B stars, starspots imply the presence of magnetic fields in stars 
with radiative envelopes.  Flares are also seen in A and late-B stars 
\citep{Balona2013c, Balona2015a, Balona2020p}.  These cannot be due to 
late-type companions since the flare energies are well beyond the largest ever 
seen in cool flare stars.  Rotational modulation and flares provide indirect 
evidence of the presence of magnetic fields in a sizeable fraction of A and B 
stars, including Be stars.

Since it is now clear that NRP cannot provide the required mass loss
mechanism, further progress must rest on development of the impulsive
magnetic rotator model or some other exploration of alternative mass-loss 
mechanisms.

\section*{Acknowledgments}

LAB wishes to thank the National Research Foundation of South Africa for 
financial support. Funding for the {\it TESS} mission is provided by the NASA 
Explorer Program. Funding for the {\it TESS} Asteroseismic Science Operations 
Centre is provided by the Danish National Research Foundation (Grant agreement 
no.: DNRF106), ESA PRODEX (PEA 4000119301) and Stellar Astrophysics Centre 
(SAC) at Aarhus University. 

This work has made use of the {\em BeSS} database, operated at LESIA, 
Observatoire de Meudon, France: http://basebe.obspm.fr.

This work has made use of data from the European Space Agency (ESA) mission 
Gaia, processed by the Gaia Data Processing and Analysis Consortium (DPAC).
Funding for the DPAC has been provided by national institutions, in particular 
the institutions participating in the Gaia Multilateral Agreement.  

This research has made use of the SIMBAD database, operated at CDS, 
Strasbourg, France.  Data were obtained from the Mikulski Archive for Space 
Telescopes (MAST).  STScI is operated by the Association of Universities for 
Research in Astronomy, Inc., under NASA contract NAS5-2655.

\bibliographystyle{mnras}
\bibliography{be}

\label{lastpage} 

\end{document}